\documentclass[11pt]{article}

\usepackage[preprint]{acl}

\usepackage{times}
\usepackage{latexsym}

\usepackage[T1]{fontenc}

\usepackage[utf8]{inputenc}

\usepackage{microtype}

\usepackage{inconsolata}

\usepackage{graphicx}
\usepackage{booktabs}
\usepackage{amsmath}
\usepackage{amssymb}
\usepackage{float}

\title{Listwise Cross-Encoder Fine-Tuning vs. Agentic Instruction Tuning for LLM Rerankers: A Systematic Study in Medical Procedure Reranking}

\author{Matan Fainzilber, Shlomit Plavner \\ Healthee \\ \texttt{\{matan.fainzilber, shlomit.plavner\}@healthee.co}}

\begin{document}
\maketitle
\begin{abstract}
Reranking medical procedures against patient queries is a critical component of health-insurance information retrieval, complicated by a substantial lexical gap between patient language and clinical nomenclature. We present a systematic comparison of two reranking paradigms for this production task: (1)~small cross-encoders (MedCPT, MiniLM-L12) fine-tuned with listwise learning-to-rank objectives across layer-freezing configurations, and (2)~Qwen3-Reranker-4B, a 4B-parameter instruction reranker whose prompt is iteratively refined via an agentic optimization loop driven by GPT-4.1. On a purpose-built dataset of 2,647 queries across 708 insurance services, we find that a 109M-parameter cross-encoder fine-tuned with ListNet outperforms the 4B-parameter model by 2.6 percentage points on NDCG@3 and 13.3 points on Spearman correlation - at 37$\times$ fewer parameters. We report practical findings, a scalable LLM-based dataset construction pipeline, and deployment trade-offs relevant to production reranking systems. We release our code and a sample dataset to support reproducibility and adaptation to other domains.\footnote{\href{https://github.com/matanf-healthee/listwise-crossencoder-reranking}{Code repository}}
\end{abstract}

\section{Introduction}

When a patient asks whether a particular medical service is covered by their insurance plan, the system must identify which billable procedures correspond to their query. In a typical production pipeline, a first-stage retriever returns a set of candidate procedures, and a reranker then scores and orders them by clinical relevance. This alignment task is complicated by a significant \textit{lexical gap}: patients describe needs using layman terms and symptom descriptions (e.g., ``my knee hurts when I walk''), while procedures are catalogued in clinical nomenclature (e.g., ``arthrocentesis of the knee joint''). Bridging this gap accurately is critical, misalignment leads to incorrect coverage determinations, claim denials, and downstream administrative burden \citep{Jin2023, Alsentzer2019}.

Recent work has demonstrated that large language models (LLMs) can serve as effective zero-shot rerankers \citep{Sun2023, Ma2023}. However, deploying billion-parameter models for real-time medical information retrieval raises practical concerns: inference latency, hosting cost, and the absence of domain-specific calibration. An alternative paradigm is to fine-tune smaller cross-encoder models on domain data, but prior work has largely relied on pointwise or pairwise objectives that do not directly optimize ranking quality across the full candidate list.

In this work, we investigate whether listwise learning-to-rank objectives can close the gap between small specialized models and large generalist rerankers in the medical insurance domain. Our contributions are as follows:

\begin{enumerate}
    \item We construct a purpose-built evaluation dataset for medical procedure reranking using a two-phase LLM synthesis pipeline with quality filtering, yielding 2,647 graded query-procedure lists across 708 insurance services.
    \item We conduct a systematic comparison of three listwise loss functions: LambdaLoss, ListNet, and PListMLE across two cross-encoder backbones and three layer-freezing strategies, comprising 18 experimental configurations.
    \item We propose an agentic prompt optimization framework for instruction-tuned rerankers that uses GPT-4.1 to iteratively refine the task instruction for Qwen3-Reranker-4B, providing a strong LLM baseline.
    \item \textbf{Practical findings}: evidence that domain-specialized cross-encoders fine-tuned with listwise objectives can achieve competitive or superior ranking quality to instruction-tuned models 37$\times$ their size, with deployment advantages in cost and latency (\S\ref{sec:results}).
\end{enumerate}

\section{Related Work}

Cross-encoders jointly encode query-document pairs to produce relevance scores \citep{Devlin2019}, outperforming bi-encoders on reranking benchmarks. Domain-adapted variants such as MedCPT \citep{Jin2023} and Clinical BERT \citep{Alsentzer2019} have demonstrated the value of in-domain pre-training for biomedical retrieval. Recently, instruction-tuned LLMs have emerged as zero-shot rerankers \citep{Sun2023, Ma2023}, but face challenges in latency, cost, and domain specificity. On the training objective side, listwise losses like LambdaLoss \citep{Wang2018}, ListNet \citep{Cao2007}, and PListMLE \citep{Lan2014} operate on entire candidate lists, with theoretical analysis connecting these losses to ranking measures such as NDCG \citep{Chen2009}. Recent work has further explored cross-encoder fine-tuning strategies for reranking \citep{Pezzuti2025}. Our work bridges these threads: we systematically compare listwise-trained small encoders against an agentically optimized LLM reranker on a real production task, reporting practical findings relevant to deployment.

\section{Task and Dataset}
\label{sec:dataset}

\subsection{Task Definition}

Given a user query $q$ describing a medical need and a set of $n$ candidate procedures $\mathcal{P} = \{p_1, \ldots, p_n\}$, the task is to learn a scoring function $f(q, p_i) \rightarrow \mathbb{R}$ that induces a ranking $\pi$ over $\mathcal{P}$ such that procedures most clinically relevant to $q$ are ranked highest. Formally, the goal is to maximize $\text{NDCG@}k$ with respect to a ground-truth relevance assignment $y_i \in \mathbb{R}_{\geq 0}$ for each $p_i$:
\begin{align}
\text{DCG@}k &= \sum_{i=1}^{k} \frac{2^{y_{\pi(i)}} - 1}{\log_2(i+1)} \label{eq:dcg} \\
\text{NDCG@}k &= \frac{\text{DCG@}k}{\text{IDCG@}k} \label{eq:ndcg}
\end{align}
where IDCG@$k$ is the DCG of the ideal (ground-truth) ranking, normalizing scores to $[0, 1]$.
Queries span three intent types: direct procedure requests (``I need an MRI''), symptom descriptions (``my knee hurts when I walk''), and coverage inquiries (``does my plan cover physical therapy?'').

\subsection{Dataset Construction}

We construct a purpose-built dataset using a two-phase LLM synthesis pipeline. The source data consists of 708 insurance services, each mapped to a subset of 1,517 unique medical procedures by domain experts.

\paragraph{Phase 1: Query Generation.}
For each service, we generate $2n$ queries (where $n$ is the number of associated procedures) using GPT-4o with temperature 0.7. Each query is assigned a target procedure via round-robin allocation over a shuffled procedure index, ensuring every procedure serves as the primary target at least twice. Queries are diversified along three independently sampled dimensions: intent type (coverage / symptom / direct request), formality (rephrased keyword-style at 70\% / conversational at 30\%), and synonym usage. The generation prompt explicitly prohibits copying procedure names verbatim, forcing the model to produce realistic layman or clinical paraphrases.

\paragraph{Phase 2: Relevance Ranking.}
A separate GPT-4o call (temperature 0.0) ranks all candidate procedures for each generated query. For services with more than 20 procedures, we subsample to the target procedure plus 19 random distractors to keep the ranking prompt tractable.

\paragraph{Quality Filtering.}
We retain only rows where the teacher model places the target procedure within the top 3 positions, discarding queries where Phase~2 could not tie the generated query back to its intended target. This yields a final dataset of 2,647 query-procedure lists. The full pipeline is formalized as pseudocode in Appendix~\ref{sec:appendix-algorithm}.

\paragraph{Human Validation.}
The underlying service-to-procedure mappings were constructed by domain experts with knowledge of US health-insurance billing. To validate the LLM-generated rankings, two domain experts independently evaluated a random sample of 100 query--procedure rankings from the final dataset, judging whether the teacher model's top-1 and top-3 placements were clinically appropriate. Expert~A accepted the top-1 placement in 92\% of cases, Expert~B in 97\%. Top-3 acceptance rates were 92\% and 97\% respectively. Raw inter-annotator agreement was 95\% (Cohen's $\kappa = 0.52$; the moderate kappa reflects the well-documented prevalence bias in highly skewed distributions). These results confirm that the GPT-4o rankings align well with expert clinical judgment, supporting the quality of the synthetic labels for training and evaluation.

\subsection{Data Splits}

We partition the dataset at the \textit{query level} to prevent data leakage: 75\% train, 15\% validation, and 10\% test. The same candidate procedure may appear under different queries across splits, but no query appears in more than one split.

\section{Methodology}
\label{sec:methodology}

\subsection{Cross-Encoder Fine-Tuning}

We fine-tune two cross-encoder backbones: \textbf{MedCPT-Cross-Encoder} \citep{Jin2023}, a biomedical model pre-trained on PubMed search logs providing strong domain initialization, and \textbf{MiniLM-L-12-v2}, a general-purpose cross-encoder distilled from MS MARCO. Both models encode a (query, procedure) pair jointly and output a single relevance logit. During training, each batch element is a full ranked list of candidates for one query, and the loss operates over the entire list.

\subsection{Listwise Loss Functions}

We evaluate three listwise objectives over raw model logits $s_i = f(q, p_i)$ with ground-truth relevance labels $y_i$:

\paragraph{LambdaLoss} \citep{Wang2018} weights pairwise gradients by NDCG gain differences:
\begin{equation}
\mathcal{L}_{\text{Lambda}} = \sum_{i} \sum_{j: y_i > y_j} |\Delta \text{NDCG}_{ij}| \cdot \log\!\big(1 + e^{-(s_i - s_j)}\big)
\label{eq:lambdaloss}
\end{equation}

\paragraph{ListNet} \citep{Cao2007} minimizes cross-entropy between ground-truth and predicted softmax ranking distributions:
\begin{equation}
\mathcal{L}_{\text{ListNet}} = -\sum_{i} P_y(p_i) \log P_s(p_i)
\label{eq:listnet}
\end{equation}

\paragraph{PListMLE} \citep{Lan2014} models the ranking as a position-weighted Plackett--Luce process:
\begin{equation}
\mathcal{L}_{\text{PListMLE}} = -\sum_{i} w_i \Big( s_{\pi(i)} - \log \sum_{j \geq i} \exp(s_{\pi(j)}) \Big)
\label{eq:plistmle}
\end{equation}

\subsection{Layer-Freezing Configurations}

We explore three strategies: \textbf{no freeze} (all parameters trainable), \textbf{freeze embeddings} (freeze token/position/type embeddings; ${\sim}78\%$ trainable), and \textbf{freeze 6 layers} (freeze embeddings + first 6 transformer layers; ${\sim}40\%$ trainable). Combined with 2 models and 3 losses, this yields an 18-cell experiment grid.

\subsection{Agentic Prompt Optimization}
\label{sec:agentic}

To establish a strong LLM baseline, we optimize the task instruction for Qwen3-Reranker-4B \citep{Zhang2025}, a 4-billion-parameter instruction-tuned reranker that accepts an explicit natural-language instruction at inference time.

We design an iterative optimization loop driven by GPT-4.1:

\begin{enumerate}
    \item \textbf{Evaluate} the current instruction on a fixed validation subset (200 queries, seed 42) and compute per-query NDCG@3.
    \item \textbf{Select examples}: identify the 10 highest- and 10 lowest-scoring queries as good and bad exemplars.
    \item \textbf{Generate candidates}: construct a meta-prompt containing the current best instruction, a history of prior iterations, and the selected examples. GPT-4.1 proposes $k=5$ candidate instructions with rationales.
    \item \textbf{Evaluate candidates} concurrently against the validation subset.
    \item \textbf{Update}: if the best candidate improves over the current best, adopt it and reset patience, otherwise increment patience. Stop after 5 consecutive non-improving iterations or 30 total iterations.
\end{enumerate}

The final instruction is evaluated on the full validation and test sets. This procedure ensures the LLM baseline is not disadvantaged by a suboptimal prompt.  To avoid leaking domain knowledge into the baseline, we start from a deliberately domain-agnostic instruction and let the optimizer discover any helpful framing from the in-domain examples alone.  \paragraph{Initial instruction:} \textit{``Rank the candidates by relevance to the query.''}  \paragraph{Optimized instruction (after 9 iterations):} \textit{``Order the procedures from most to least relevant based on how thoroughly each aligns with the stated symptoms, diagnostic needs, or procedural requests in the query.''}  \noindent Notably, the optimizer rediscovered the medical-domain framing (``procedures'', ``symptoms'', ``diagnostic needs'') purely from the iterative signal, starting from a prompt with no domain vocabulary.

\begin{figure}[t]
\centering
\includegraphics[width=\columnwidth]{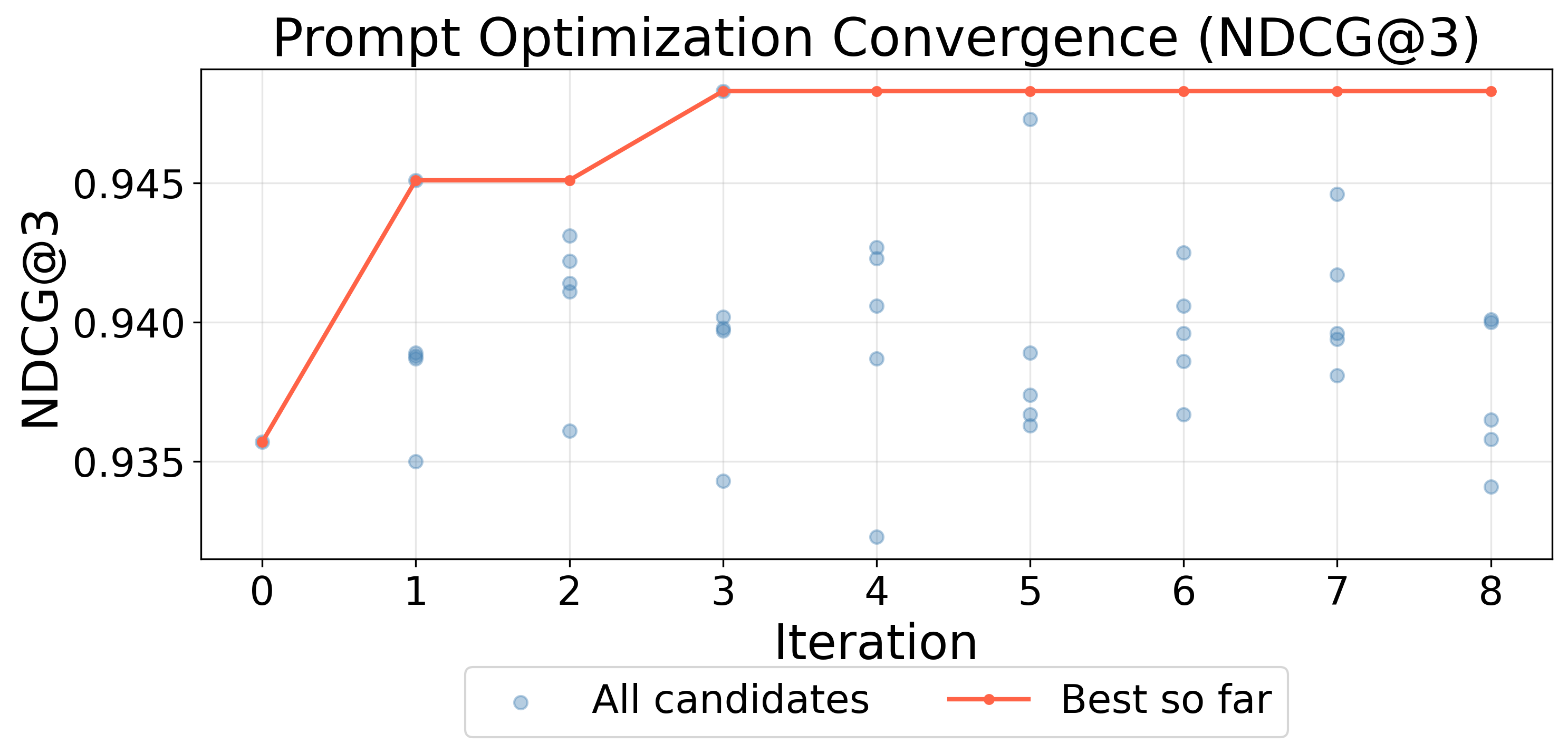}
\caption{Convergence of agentic prompt optimization for Qwen3-Reranker-4B. Blue dots: all GPT-4.1-proposed candidate instructions per iteration; red line: best-so-far NDCG@3 on the 200-query validation subsample. The optimizer converged within 3 iterations and terminated at iteration 9 via early stopping (patience 5). See Appendix~\ref{sec:appendix-tree} for a schematic of the greedy search procedure.}
\label{fig:agentic-convergence}
\end{figure}

\section{Experimental Setup}

\paragraph{Metrics.}
We report NDCG@$k$ for $k \in \{1, 3, 5\}$, Mean Reciprocal Rank (MRR), and Spearman rank correlation. The primary metric for model selection and early stopping is NDCG@3.

\paragraph{Baselines.}
Each fine-tuned model is compared against its base model checkpoint to measure the absolute gain from listwise training. All models are additionally compared against Qwen3-Reranker-4B with its agentically optimized instruction.

\paragraph{Training details.}
The initial grid search across all 18 configurations uses a uniform hyperparameter set: learning rate $2 \times 10^{-5}$ with cosine scheduling, warmup ratio 0.1, weight decay 0.01, gradient clipping 1.0, batch size 32 ranked lists, and FP16 mixed precision. Training runs for a maximum of 50 epochs with early stopping (patience 5, minimum 15 epochs). Relevance labels follow a linear scheme: for $n$ candidates, scores are assigned as $[n, n{-}1, \ldots, 1]$.

\paragraph{Evaluation protocol.}
All model selection and hyperparameter decisions are made on the validation set (400 queries). The test set (263 queries) is held out and evaluated exactly once for final reported results.

\paragraph{Hyperparameter tuning.}
Based on validation NDCG@3, we select the top 4 configurations and run Optuna (TPE sampler, 50 trials per config) over the search space in Table~\ref{tab:hp-search}, with validation NDCG@3 as the objective. The search revealed that LR scheduler choice had negligible importance, consistent with early stopping firing well before the schedule diverges. We therefore standardize the final training recipe on \texttt{constant\_with\_warmup}, patience 15, weight decay 0.01, 100-epoch ceiling, and tune per-config the three HPs whose optima differed: learning rate, batch size, and warmup ratio. Each final configuration is trained with 5 random seeds (42, 123, 456, 789, 1024), then we report mean $\pm$ std on the held-out test set.

\section{Results and Analysis}
\label{sec:results}

\subsection{Grid Search Results}

We first conduct a systematic grid search across all 18 configurations (2 models $\times$ 3 losses $\times$ 3 freeze strategies) using uniform hyperparameters. Full validation-set results are reported in Appendix~\ref{sec:appendix-grid}. All 18 fine-tuned configurations improve over their respective base models across every metric. MedCPT outperforms MiniLM in absolute terms across all configurations: all 9 MedCPT cells exceed all 9 MiniLM cells on validation NDCG@3. Ranked by validation NDCG@3, the top 4 configurations form a 2$\times$2 ablation: MedCPT with \{LambdaLoss, ListNet\}$\times$\{no\_freeze, freeze\_emb\}, with validation NDCG@3 ranging from .9611 to .9636. PListMLE and freeze\_6\_layers are eliminated based on consistently lower validation performance. These 4 configurations are selected for targeted hyperparameter optimization.

\subsection{Hyperparameter Tuning}

For each of the top 4 chosen configurations, we run 50 Optuna trials (TPE sampler) over the search space in Table~\ref{tab:hp-search}, optimizing validation NDCG@3.

\begin{table}[t]
\centering
\small
\begin{tabular}{ll}
\toprule
\textbf{Hyperparameter} & \textbf{Range} \\
\midrule
Learning rate & log-uniform $[5\!\times\!10^{-6},\; 10^{-4}]$ \\
Weight decay & log-uniform $[10^{-4},\; 10^{-1}]$ \\
Warmup ratio & uniform $[0.01,\; 0.3]$ \\
Batch size & $\{16, 32, 64\}$ \\
ES patience & $\{5, 10, 15\}$ \\
LR scheduler & $\{\text{cosine}, \text{linear}\}$ \\
\bottomrule
\end{tabular}
\caption{Optuna search space (TPE sampler, 50 trials per config, seed 42). Best values per config in Appendix~\ref{sec:appendix-hyperparams}.}
\label{tab:hp-search}
\end{table}

\subsection{Final Results}

Table~\ref{tab:final-results} presents the test-set performance of the top 4 configurations after hyperparameter tuning, alongside base models and Qwen3-Reranker-4B. Fine-tuned results are mean $\pm$ std across 5 random seeds.

\begin{table*}[t]
\centering
\small
\begin{tabular}{llll|ccccc}
\toprule
\textbf{Model} & \textbf{Loss} & \textbf{Freeze} & \textbf{Params} & \textbf{NDCG@1} & \textbf{NDCG@3} & \textbf{NDCG@5} & \textbf{MRR} & \textbf{Spearman} \\
\midrule
MedCPT (base) & -- & -- & 109M & .9135 & .9153 & .9202 & .8569 & .5845 \\
MiniLM (base) & -- & -- & 33M & .9144 & .8804 & .8854 & .8466 & .4774 \\
\midrule
MedCPT & ListNet & no\_freeze & 109M & \textbf{.957}$_{\pm.003}$ & \textbf{.961}$_{\pm.002}$ & \textbf{.963}$_{\pm.002}$ & \textbf{.924}$_{\pm.006}$ & \textbf{.779}$_{\pm.011}$ \\
MedCPT & ListNet & freeze\_emb & 109M & .956$_{\pm.004}$ & .960$_{\pm.002}$ & .962$_{\pm.003}$ & .924$_{\pm.008}$ & .770$_{\pm.013}$ \\
MedCPT & LambdaLoss & no\_freeze & 109M & .951$_{\pm.003}$ & .958$_{\pm.003}$ & .958$_{\pm.001}$ & .921$_{\pm.008}$ & .754$_{\pm.017}$ \\
MedCPT & LambdaLoss & freeze\_emb & 109M & .951$_{\pm.004}$ & .957$_{\pm.003}$ & .958$_{\pm.003}$ & .920$_{\pm.008}$ & .748$_{\pm.018}$ \\
\midrule
Qwen3-4B & -- & -- & 4B & .941 & .936 & .940 & .898 & .646 \\
\bottomrule
\end{tabular}
\caption{Test-set results (263 queries). Fine-tuned configs: mean$_{\pm\text{std}}$ across 5 seeds. Qwen3-Reranker-4B with agentically optimized prompt. Sorted by NDCG@3. Best per metric in \textbf{bold}.}
\label{tab:final-results}
\end{table*}

The best configuration, MedCPT with ListNet and no freezing, achieves 0.961 NDCG@3 ($\pm$0.002), a +4.6pp improvement over the base model MedCPT baseline. ListNet slightly outperforms LambdaLoss on the test set across both freeze settings, while the freeze axis shows minimal effect ($<$0.1pp within each loss family). The largest gains appear on Spearman correlation (+19.4pp), indicating that listwise training particularly improves full-list ordering beyond the top position. Seed standard deviations are tight ($\hat{\sigma} < 0.003$ on NDCG@3).

\subsection{Effect of Loss Functions and Layer Freezing}

\begin{figure}[t]
\centering
\includegraphics[width=\columnwidth]{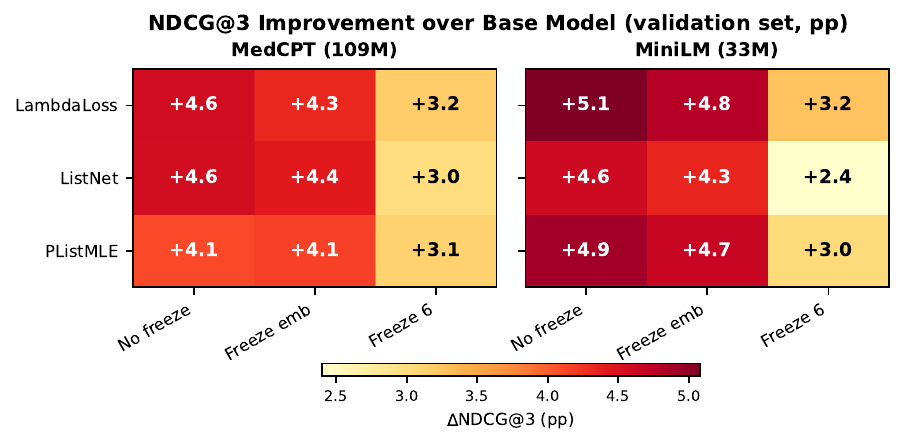}
\caption{NDCG@3 improvement (pp) over base model on the validation set across the full 18-cell grid. MiniLM shows larger relative gains but is more sensitive to aggressive freezing.}
\label{fig:heatmap}
\end{figure}

Figure~\ref{fig:heatmap} summarizes the NDCG@3 improvements on the validation set across the full grid (see Appendix~\ref{sec:appendix-grid} for all 18 configurations). Averaging across all models and freeze configurations, LambdaLoss leads on validation NDCG@1 (.9585) and NDCG@3 (.9491), followed by PListMLE (.9549, .9470) and ListNet (.9534, .9460). However, the margins between losses are small compared to the effect of freezing strategy.

For freezing, no freeze and freeze embeddings perform similarly, with no freeze holding a slight edge on NDCG@3 (.9534 vs.\ .9515) and Spearman (.7392 vs.\ .7296). Freezing 6 layers incurs a substantial drop: $-$1.6pp on NDCG@3 and $-$5.6pp on Spearman relative to no freeze. MedCPT tolerates aggressive freezing better than MiniLM, consistent with its biomedical pre-training providing more robust lower-layer representations.

All three losses yield large Spearman improvements over base models (+12-18pp for MedCPT, +10-19pp for MiniLM), confirming that listwise training substantially improves full-list rank correlation regardless of the specific objective.

\subsection{Cross-Encoder vs.\ LLM Reranker}

\begin{figure}[t]
\centering
\includegraphics[width=\columnwidth]{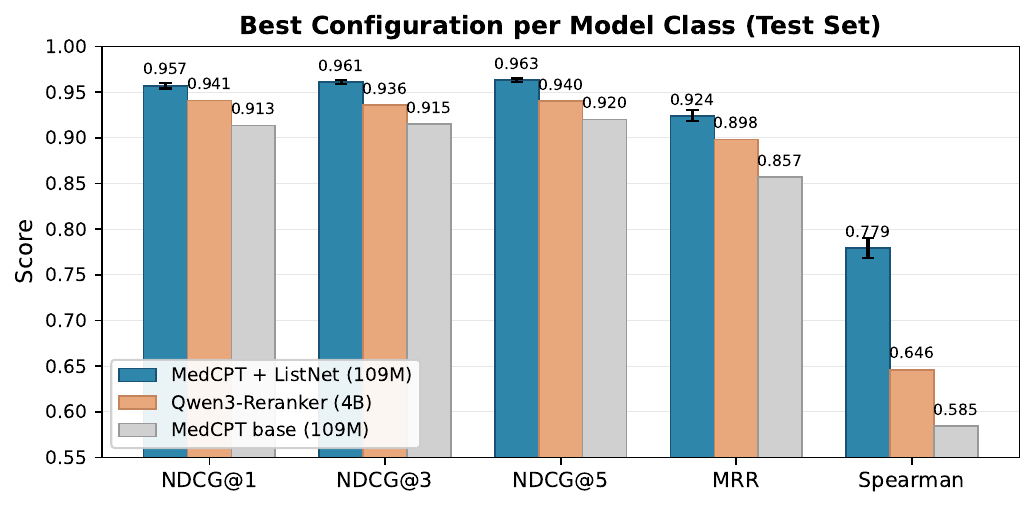}
\caption{Best fine-tuned cross-encoder (MedCPT + ListNet, 109M) vs.\ prompt-optimized Qwen3-Reranker (4B) vs.\ unfine-tuned MedCPT baseline on the held-out test set. CE error bars show $\pm$1 std across 5 seeds.}
\label{fig:specialist-generalist}
\end{figure}

Table~\ref{tab:final-results} and Figure~\ref{fig:specialist-generalist} present the complete comparison. The best fine-tuned MedCPT (ListNet, no\_freeze) outperforms the prompt-optimized 4B-parameter Qwen3-Reranker on all metrics: +1.6pp on NDCG@1, +2.6pp on NDCG@3, and +13.3pp on Spearman - while using 37$\times$ fewer parameters. These gaps substantially exceed the cross-encoder seed variance ($\sigma \approx 0.002$). Even the weakest of the four fine-tuned configurations (LambdaLoss, freeze\_emb) leads Qwen3 by +2.1pp on NDCG@3. Qwen3's Spearman correlation (0.646) is notably closer to the MedCPT baseline (0.585) than to any fine-tuned model ($\geq$0.748), suggesting that prompt optimization alone cannot match the full-list ordering calibration that listwise training provides.

\section{Discussion and Practical Takeaways}

\paragraph{Fine-tuned cross-encoders as production rerankers.}
Our best 109M-parameter cross-encoder (MedCPT + ListNet, no\_freeze) achieves 0.961 test NDCG@3 ($\pm$0.002 across 5 seeds), a +4.6pp absolute improvement over the unfine-tuned MedCPT baseline and a +19.4pp lift on Spearman correlation. Cross-encoders of this scale can be served on commodity CPU hardware, whereas 4B-parameter models require dedicated GPUs. For production systems processing thousands of queries per hour, this difference in serving cost is significant.

\paragraph{Scalable dataset construction via LLM synthesis.}
Our two-phase pipeline for query generation followed by independent ranking with quality filtering provides a practical path for constructing domain-specific reranking datasets when manual annotation is prohibitively expensive. The round-robin target assignment ensures broad procedure coverage, and the top-3 quality filter effectively removes noisy examples.

\paragraph{Agentic prompt optimization.}
The iterative optimization loop improved Qwen3-Reranker's validation NDCG@3 from 0.936 (domain-agnostic seed) to 0.948 within 3 iterations, demonstrating that instruction-tuned rerankers benefit substantially from task-specific prompts. The optimizer autonomously rediscovered medical-domain framing from in-domain exemplars alone. Despite this, the prompt-optimized 4B-parameter model still trails the best 109M-parameter fine-tuned cross-encoder by 2.6pp on test NDCG@3, suggesting that prompt-only adaptation cannot substitute for listwise fine-tuning on domain data.

\section{Conclusion}

We presented a systematic comparison of listwise-trained cross-encoders and an agentically optimized instruction reranker for medical procedure alignment in a production health-insurance system. Through an 18-cell experiment grid spanning two backbones, three listwise losses, and three layer-freezing strategies, we found that 109M-parameter MedCPT models fine-tuned with listwise objectives outperform a prompt-optimized 4B-parameter Qwen3-Reranker by 2.6pp on NDCG@3 and 13.3pp on Spearman correlation, with tight seed variance ($\hat\sigma < 0.003$), despite using 37$\times$ fewer parameters. Our key practical takeaways: LLM-based dataset synthesis provides a scalable alternative to manual annotation for domain-specific reranking, agentic prompt optimization offers a principled way to establish strong LLM baselines, and domain-specialized small models offer a compelling production alternative to large generalist rerankers when serving cost and latency are constrained.

\section*{Limitations}

Our study has several limitations. First, the dataset is constructed from a single organization's service-to-procedure mappings and may not generalize to other insurance taxonomies or international healthcare systems. Second, the relevance rankings used for training are produced by GPT-4o acting as a teacher model: any systematic biases of this teacher are inherited by the training labels, although our top-3 quality filter and human validation (\S\ref{sec:dataset}) partially mitigate this. Third, we evaluate only two cross-encoder backbones and one instruction reranker. A broader comparison including models such as GTE, BGE, or larger instruction rerankers would strengthen the conclusions. Our freezing strategies also did not explore the reverse configuration, freezing upper layers while keeping embeddings trainable, which may be particularly relevant for lexical-gap tasks where adapting the token representations could matter more than modifying higher-level reasoning layers.

\section*{Ethics Statement}

This research uses de-identified internal metadata consisting of service names, procedure descriptions, and synthetically generated queries. No patient data, protected health information, or personally identifiable information was used at any stage. The models described in this work are intended exclusively for administrative decision support, aligning procedure codes with insurance coverage queries, and are not designed or validated for clinical diagnosis, treatment recommendation, or any direct patient-facing medical decision. The domain experts involved in dataset construction and validation were salaried professionals with standard compensation. Our data synthesis pipeline relies on GPT-4o, and we acknowledge that synthetic data may inherit linguistic or demographic biases from the teacher model, potentially affecting performance across diverse patient phrasings. Finally, while our offline experiments incurred usage of GPU compute, our core finding, that 109M-parameter cross-encoders outperform 4B-parameter models, directly supports reduced inference cost and energy consumption in deployment.

\bibliography{custom}

\appendix
\newpage
\onecolumn

\section{Full Grid Search Results}
\label{sec:appendix-grid}

Table~\ref{tab:grid-results} presents the complete 18-cell grid search results on the validation set (400 queries) using uniform hyperparameters. These results informed the selection of the top 4 configurations for targeted hyperparameter tuning.

\begin{table}[H]
\centering
\small
\begin{tabular}{llll|ccccc}
\toprule
\textbf{Model} & \textbf{Loss} & \textbf{Freeze} & \textbf{Params} & \textbf{NDCG@1} & \textbf{NDCG@3} & \textbf{NDCG@5} & \textbf{MRR} & \textbf{Spearman} \\
\midrule
MedCPT (base) & -- & -- & 109M & .9265 & .9179 & .9222 & .8732 & .5995 \\
MedCPT & LambdaLoss & no\_freeze & 109M & \textbf{.9702} & \textbf{.9636} & \textbf{.9622} & \textbf{.9427} & \textbf{.7785} \\
MedCPT & LambdaLoss & freeze\_emb & 109M & .9662 & .9611 & .9606 & .9360 & .7721 \\
MedCPT & LambdaLoss & freeze\_6 & 109M & .9601 & .9497 & .9523 & .9193 & .7364 \\
MedCPT & ListNet & no\_freeze & 109M & .9697 & .9635 & .9612 & .9418 & .7695 \\
MedCPT & ListNet & freeze\_emb & 109M & .9694 & .9621 & .9607 & .9373 & .7624 \\
MedCPT & ListNet & freeze\_6 & 109M & .9583 & .9479 & .9491 & .9222 & .7145 \\
MedCPT & PListMLE & no\_freeze & 109M & .9635 & .9589 & .9573 & .9257 & .7657 \\
MedCPT & PListMLE & freeze\_emb & 109M & .9626 & .9594 & .9572 & .9274 & .7648 \\
MedCPT & PListMLE & freeze\_6 & 109M & .9605 & .9492 & .9501 & .9210 & .7167 \\
\midrule
MiniLM (base) & -- & -- & 33M & .9308 & .8963 & .8964 & .8691 & .5314 \\
MiniLM & LambdaLoss & no\_freeze & 33M & .9570 & .9471 & .9476 & .9099 & .7190 \\
MiniLM & LambdaLoss & freeze\_emb & 33M & .9552 & .9440 & .9452 & .9094 & .7031 \\
MiniLM & LambdaLoss & freeze\_6 & 33M & .9424 & .9288 & .9301 & .8835 & .6549 \\
MiniLM & ListNet & no\_freeze & 33M & .9462 & .9423 & .9412 & .9016 & .6964 \\
MiniLM & ListNet & freeze\_emb & 33M & .9428 & .9397 & .9392 & .9021 & .6904 \\
MiniLM & ListNet & freeze\_6 & 33M & .9340 & .9203 & .9225 & .8838 & .6270 \\
MiniLM & PListMLE & no\_freeze & 33M & .9529 & .9452 & .9442 & .9013 & .7061 \\
MiniLM & PListMLE & freeze\_emb & 33M & .9484 & .9429 & .9410 & .8943 & .6847 \\
MiniLM & PListMLE & freeze\_6 & 33M & .9417 & .9266 & .9271 & .8876 & .6489 \\
\bottomrule
\end{tabular}
\caption{Full grid search results on the validation set (400 queries, uniform hyperparameters). Sorted by model, then loss, then freeze config. Best per metric in \textbf{bold}. The top 4 configs (all MedCPT, NDCG@3 $\geq$ .9611) were selected for Optuna hyperparameter tuning.}
\label{tab:grid-results}
\end{table}

\section{Hyperparameter Details}
\label{sec:appendix-hyperparams}

Table~\ref{tab:hp-best} reports the per-config hyperparameters selected by Optuna and used for the final multi-seed training. Standardized settings across all configs: \texttt{constant\_with\_warmup} scheduler, weight decay 0.01, patience 15, 100-epoch ceiling, FP16.

\begin{table}[H]
\centering
\small
\begin{tabular}{llll}
\toprule
\textbf{Config} & \textbf{LR} & \textbf{BS} & \textbf{Warmup} \\
\midrule
$\lambda$Loss + no\_freeze & 6.08e-5 & 16 & 0.099 \\
$\lambda$Loss + freeze\_emb & 7.25e-5 & 32 & 0.066 \\
ListNet + no\_freeze & 1.71e-5 & 16 & 0.204 \\
ListNet + freeze\_emb & 5.53e-5 & 32 & 0.173 \\
\bottomrule
\end{tabular}
\caption{Per-config hyperparameters from Optuna (50 trials each, TPE sampler). All configs use MedCPT backbone.}
\label{tab:hp-best}
\end{table}

\section{Additional Results}
\label{sec:appendix-results}

\begin{figure}[H]
\centering
\includegraphics[width=0.65\textwidth]{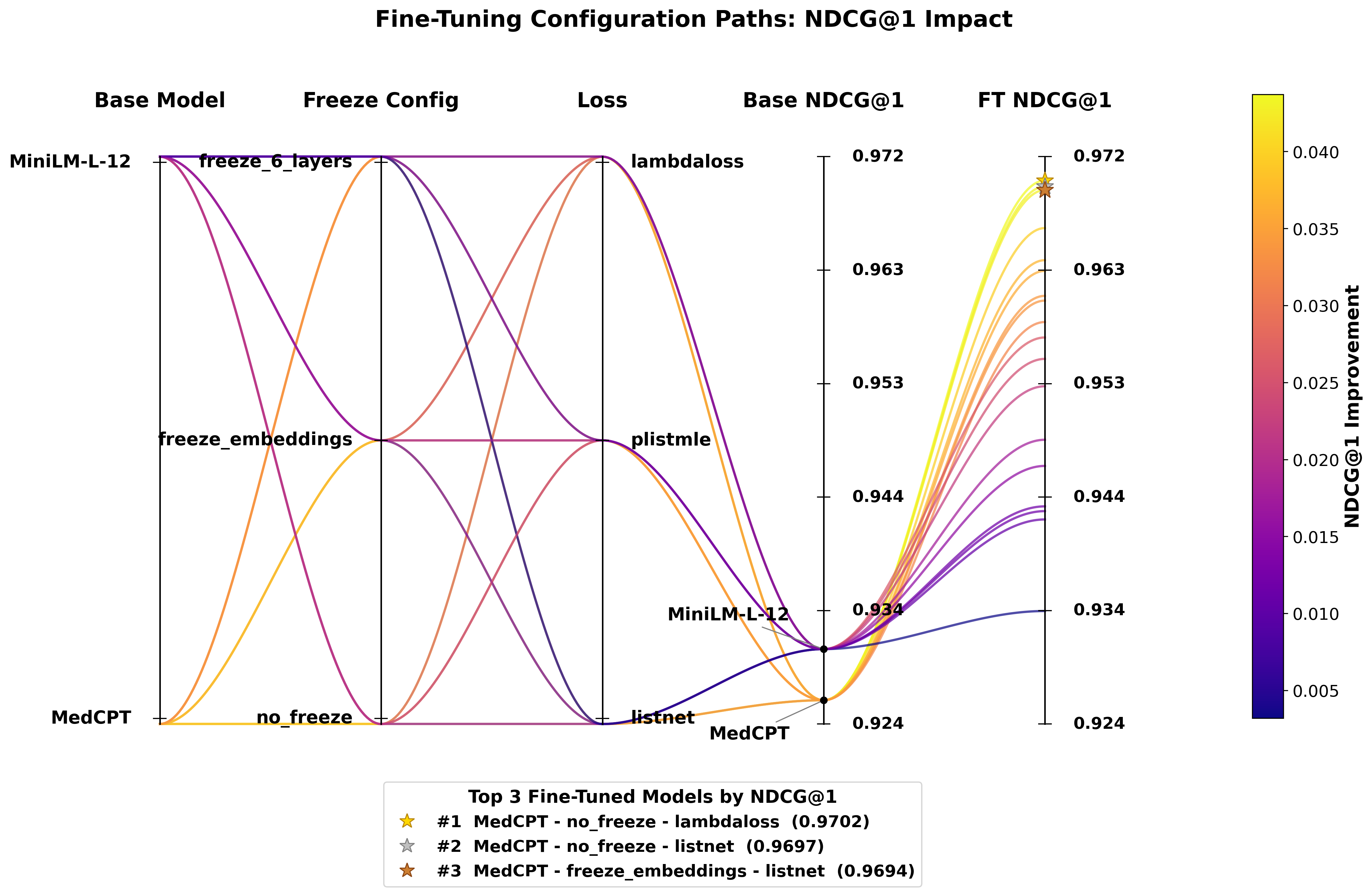}
\caption{Parallel coordinates plot of all 18 fine-tuning configurations on the validation set. Each line traces a configuration from base model to final NDCG@1, colored by improvement magnitude. Stars mark the top 3 configurations.}
\label{fig:config-paths}
\end{figure}

Figure~\ref{fig:config-paths} visualizes the full experiment grid as a parallel coordinates plot on the validation set, tracing each of the 18 fine-tuned configurations from base model through freeze strategy and loss function to final NDCG@1. Line color encodes the NDCG@1 improvement over the base model.

\section{Agentic Search Procedure}
\label{sec:appendix-tree}

\begin{figure}[H]
\centering
\includegraphics[width=0.5\textwidth]{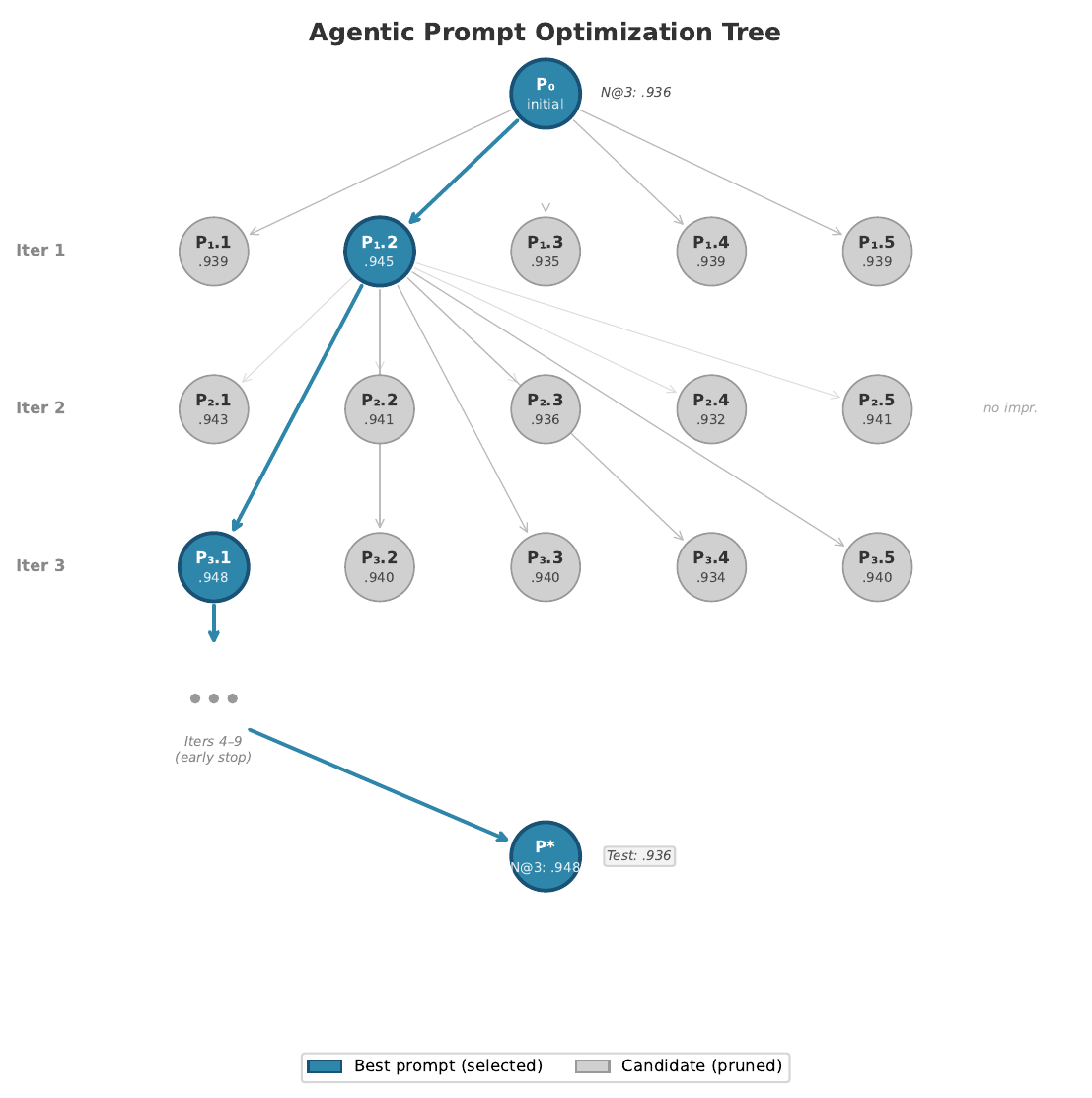}
\caption{Schematic of the agentic prompt optimization procedure as a greedy tree search. Blue nodes: selected best prompt per iteration; grey nodes: pruned candidates. Scores are NDCG@3 on the validation subsample.}
\label{fig:agentic-tree}
\end{figure}

Figure~\ref{fig:agentic-tree} illustrates the greedy tree search procedure used for agentic prompt optimization. At each iteration, the optimizer generates $k$ candidate instructions from the current best; only the highest-scoring candidate is retained and expanded in the next iteration. When no candidate improves over the current best (e.g., Iter~2), the next iteration's arrows trace back to the last improving node.

\section{Dataset Construction Algorithm}
\label{sec:appendix-algorithm}

Figure~\ref{alg:dataset} formalizes the two-phase LLM synthesis pipeline described in \S\ref{sec:dataset}. For each service, combinations are constructed via round-robin target assignment with independent scenario sampling, then processed through query generation (Phase~1), relevance ranking (Phase~2), and a quality gate.

\begin{figure}[H]
\small
\begin{tabular}{p{\textwidth}}
\toprule
\textbf{Dataset Construction Pipeline} \\
\midrule
\textbf{Input:} Service catalogue $\mathcal{S}$; each $s \in \mathcal{S}$ maps to procedures $\mathcal{P}_s$ \\
\textbf{Output:} Dataset $\mathcal{D} = \{(q_i, \pi_i)\}$ of query--ranking pairs for cross-encoder training \\[4pt]

$\mathcal{D} \leftarrow \emptyset$ \\
\textbf{for each} service $s \in \mathcal{S}$ \textbf{do} \\
\quad \textbf{if} $|\mathcal{P}_s| < 2$ \textbf{then skip} \\
\quad $n_q \leftarrow 2 \times |\mathcal{P}_s|$ \hfill \textit{// query budget} \\
\quad $\textit{targets} \leftarrow \text{round-robin}(\text{shuffle}(\mathcal{P}_s),\; n_q)$ \\[3pt]

\quad \textbf{for} $i = 1, \ldots, n_q$ \textbf{do} \\
\quad\quad Sample intent $\in \{\text{coverage, symptom, direct}\}$ \\
\quad\quad Sample formality $\in \{\text{keyword}_{70\%},\; \text{conversational}_{30\%}\}$ \\
\quad\quad Sample synonym usage $\in \{\text{true}_{30\%},\; \text{false}_{70\%}\}$ \\[3pt]

\quad\quad \textit{// Candidate list (subsample if large)} \\
\quad\quad \textbf{if} $|\mathcal{P}_s| > 20$ \textbf{then} $C_i \leftarrow \{\textit{targets}[i]\} \cup \text{sample}(\mathcal{P}_s \setminus \{\textit{targets}[i]\},\; 19)$ \quad \textbf{else} $C_i \leftarrow \mathcal{P}_s$ \\[3pt]

\quad\quad \textit{// Phase 1: Query generation} \\
\quad\quad $q_i \leftarrow \text{GPT-4o}(s,\; C_i,\; \textit{targets}[i],\; \text{intent},\; \text{formality},\; \text{synonym flag})$ \hfill temperature $= 0.7$; no verbatim procedure names \\[3pt]

\quad\quad \textit{// Phase 2: Relevance ranking} \\
\quad\quad $\pi_i \leftarrow \text{GPT-4o}(q_i,\; C_i)$ \hfill temperature $= 0.0$ \\
\quad\quad $r_i \leftarrow \text{rank}(\textit{targets}[i],\; \pi_i)$ \\[3pt]

\quad\quad \textit{// Quality gate: retain only if target procedure ranked in top 3} \\
\quad\quad \textbf{if} $r_i \leq 3$ \textbf{then} $\mathcal{D} \leftarrow \mathcal{D} \cup \{(q_i,\; \pi_i)\}$ \\[2pt]

\textbf{return} $\mathcal{D}$ \hfill $|\mathcal{D}| = 2{,}647$ after filtering \\
\bottomrule
\end{tabular}
\caption{Dataset construction pipeline pseudocode. Each entry $(q_i, \pi_i)$ pairs a generated query with a full relevance-ranked procedure list.}
\label{alg:dataset}
\end{figure}

\end{document}